\documentclass{bmcart}

\usepackage[utf8]{inputenc} 

\def\includegraphics{}

\usepackage{xspace}
\usepackage{multirow, makecell}
\usepackage{todonotes}
\usepackage{algorithm}
\usepackage{amsmath}
\usepackage{comment}
\usepackage{url}
\usepackage{ulem}
\startlocaldefs
\endlocaldefs

\begin{document}
\newcommand{\name}{{\sc \textbf{Lerna}}\xspace}
\newcommand{\ie}{{\textit{i.e.,}}\xspace}
\newcommand{\eg}{{\textit{e.g.,}}\xspace}

\newcommand{\chaterji}[1]{\todo[inline,color=blue!40!red]{Chaterji: #1}}
\newcommand{\atul}[1]{\todo[inline,color=yellow]{Atul: #1}}
\newcommand{\pranjal}[1]{\todo[inline,color=orange]{Pranjal: #1}}
\newcommand{\kanak}[1]{\todo[inline,color=green]{Kanak: #1}}
\newcommand{\AMcomment}[1]{\todo[inline,color=red!40]{Ashraf: #1}}

\iffalse
\newcommand{\rev}[1]{\textcolor{blue}{#1}}
\newcommand{\revremoved}[1]{\sout{#1}}
\newcommand{\SCcomment}[1]{\todo[inline,color=green!40]{Somali: #1}}
\else
\newcommand{\rev}[1]{{#1}}
\newcommand{\revremoved}[1]{}
\newcommand{\SCcomment}[1]{{}}
\fi

\iffalse
\newcommand{\revv}[1]{\textcolor{purple}{#1}}
\newcommand\revvremoved{\bgroup\markoverwith{\textcolor{purple}{\rule[0.5ex]{2pt}{0.4pt}}}\ULon}
\else
\newcommand{\revv}[1]{{#1}}
\newcommand\revvremoved[1]{}
\fi

\begin{frontmatter}

\begin{fmbox}
\dochead{Research}


\title{Lerna: Transformer Architectures for Configuring Error Correction Tools for Short- and 
Long-Read Genome Sequencing}

%

\author[
   addressref={aff1},
   noteref={n1},
   email={sharm438@purdue.edu}
]{\inits{AS}\fnm{Atul} \snm{Sharma}}
\author[
   addressref={aff2},
   noteref={n1},
   email={pranjal0000@gmail.com}
]{\inits{PJ}\fnm{Pranjal} \snm{Jain}}
\author[
   addressref={aff1},
   email={amahgoub@purdue.edu}
]{\inits{AM}\fnm{Ashraf} \snm{Mahgoub}}
\author[
   addressref={aff1},
   email={zhou1248@purdue.edu}
]{\inits{ZZ}\fnm{Zihan} \snm{Zhou}}
\author[
   addressref={aff3},
   email={mahadik@adobe.com}
]{\inits{KM}\fnm{Kanak} \snm{Mahadik}}
\author[
   addressref={aff1},
   corref={aff1},
   email={schaterji@purdue.edu}
]{\inits{SC}\fnm{Somali} \snm{Chaterji}}


\address[id=aff1]{
  \orgname{Purdue University}, 
  \city{West Lafayette},                              
  \cny{US}                                    
}
\address[id=aff2]{%
  \orgname{Indian Institute of Technology Bombay},
  \city{Mumbai},
  \cny{India}
}
\address[id=aff3]{%
  \orgname{Adobe Research},
  \city{San Jose},
  \cny{US}
}

\begin{artnotes}
\note[id=n1]{Equal contributor} 
\end{artnotes}

\end{fmbox}


\begin{abstractbox}

\begin{abstract} 
\parttitle{Background} 
Sequencing technologies are prone to errors, making error correction (EC) necessary for downstream applications. EC tools need to be manually configured for optimal performance. We find that the optimal parameters (\eg $k$-mer size) are both tool- and dataset-dependent. Moreover, evaluating the performance (\ie Alignment-rate or Gain) of a given tool usually relies on a reference genome, but quality reference genomes are not always available. We introduce \name for the automated configuration of $k$-mer-based EC tools. \name first creates a language model (LM) of the uncorrected genomic reads, and then, based on this LM, calculates a metric called the \textit{perplexity metric} to evaluate the corrected reads for different parameter choices. Next, it finds the one that produces the highest alignment rate \textit{without} using a reference genome. The fundamental intuition of our approach is that the perplexity metric is inversely correlated with the quality of the assembly after error correction. Therefore, \name leverages the perplexity metric for automated tuning of $k$-mer sizes without needing a reference genome.

\parttitle{Results}
First, we show that the best $k$-mer value can vary for different datasets, even for the same EC tool. This motivates our design that automates $k$-mer size selection without using a reference genome. Second, we show the gains of our LM using its component attention-based transformers. We show the model's estimation of the perplexity metric before and after error correction. The lower the perplexity after correction, the better the $k$-mer size. We also show that the alignment rate and assembly quality computed for the corrected reads are strongly \textit{negatively} correlated with the perplexity, enabling the automated selection of $k$-mer values for better error correction, and hence, improved assembly quality. We validate our approach on both short and long reads. Additionally, we show that our attention-based models have significant runtime improvement for the entire pipeline --- 18$\times$ faster than previous works, due to parallelizing the attention mechanism and the use of JIT compilation for GPU inferencing.

\parttitle{Conclusion} 
Lerna improves \textit{de novo} genome assembly by optimizing EC tools. Our code is made available in a public repository at:
\url{https://github.com/icanforce/lerna-genomics}.
\end{abstract}




\end{abstractbox}
%

\end{frontmatter}



\section*{Background}


The drop in high-throughput sequencing costs has offered unprecedented opportunities to characterize genomes, metagenomes, and single-cell genomes across the tree-of-life.
Third-generation sequencing technologies~\cite{biosciences2015detecting,eisenstein2012oxford} have demonstrated the potential to 
produce unparalleled genome assemblies due to their capability to generate staggeringly longer reads, at a much faster pace, and remarkably low costs~\cite{schadt2010window, mahadik2019scalable}, albeit, at low signal-to-noise ratios. This fundamental challenge makes genomic sequence identification and assembly, challenging, often necessitating hybrid correction approaches over self-correction 
from a consensus of long reads~\cite{fu2019comparative}. 
\rev{The error rates in base calling in these third-generation reads are nearly two orders of magnitude higher than their second-generation counterparts~\cite{laehnemann2016denoising}, with PacBio and Oxford Nanopore reaching error rates of 15\% and 40\%~\cite{korlach2013understanding, laver2015assessing} respectively, albeit improving with more sophisticated instrumentation~\cite{amarasinghe2020opportunities}. In order to use long-read datasets, especially the more erroneous ones from the earlier versions of the PacBio and Nanopore sequencers, error correction tools in the assembly pipeline are needed~\cite{berlin2015assembling}. A majority of already existing datasets have a significant error rate in them and performing error correction on these datasets reduces the cost compared to collecting the reads again with the new technology.}
Hence, diverse EC tools have been developed to remedy the erroneous reads~\cite{salmela2014lordec}.

\subsection*{$k$-mer based EC tools}
The most dominant technique used by EC tools is $k$-mer-based error correction~\cite{yang2010reptile,liu2013musket,heo2014bless,benoit2014bloocoo,lim2014trowel}. In these $k$-spectrum-based methods, the goal is to convert insolid $k$-mers (\ie, those likely to be erroneous) to solid ones (\ie, those likely to be correct). The $k$-mers that appear above a certain threshold frequency, and are therefore expected to be legitimate, are solid $k$-mers, the others are called insolid or untrusted $k$-mers.
It has been found that performance of EC tools is extremely sensitive to the chosen  $k$-value~\cite{akogwu2016comparative, mahadik2019scalable}, and the optimal value is both tool- and dataset-dependent. Small values of $k$ result in an increase in the probability of overlap between reads at the cost of not allowing the algorithm to distinguish between erroneous and correct $k$-mers. In contrast, large $k$-values decrease the overlap probability and most $k$-mers appear to be unique, thus reducing the likelihood of correcting the errors. 

Popular EC tools such as  Lighter~\cite{song2014lighter} and LoRDEC~\cite{salmela2014lordec} for short- (\textless 400 base pairs) and long-read sequences (\textgreater 400 base pairs) respectively, require the user to select the $k$-value. Determining a favorable $k$-value among possible ones 
has been explicitly pointed out as an open area of work~\cite{abdallah2019athena, kao2011echo, mahadik2019scalable}, since an arbitrary $k$-value could generate sub-optimal assemblies. 
In these scenarios, the best $k$-values need to be found by exploring all the possible $k$-values ~\cite{kao2011echo}.
A large number of EC tools rely on $k$-mers to make corrections~\cite{yang2010reptile,song2014lighter,liu2013musket,heo2014bless,benoit2014bloocoo,lim2014trowel}. We have used popular $k$-mer-based EC tools---Lighter~\cite{song2014lighter} and LoRDEC~\cite{salmela2014lordec} for short- and long-reads error correction respectively. Both of these exemplar tools require the user to select the $k$-value. Performance of error correction is extremely sensitive to this value. 
Some existing tools (\eg KMERGENIE~\cite{chikhi2013informed}) provide intuitive heuristics 
(\eg abundance histograms)
to guide $k$-value selection when performing de Bruijn Graph (DBG)-based genome assembly. However, 
experiments have shown that the optimum value of $k$ varies across datasets \textit{and} across different EC tools~\cite{abdallah2019athena}. Thus, there is a need for a {\textit{data-driven and tool-specific}} method to select the optimal $k$-value.

The runtimes of alignment algorithms like Bowtie2~\cite{langmead2012fast} have superlinear complexity~\cite{baichoo2017computational}. The perplexity computation in \name,
in contrast, is linear in the length of the input (number of reads $\times$ read length). Because \name does not need to perform the alignment step, the runtime is significantly reduced. Furthermore, because \name subsamples reads, unlike alignment that requires entire read datasets, the time is further reduced. As seen in Table~\ref{tb1:Lighter-Blue-Racer-Perplexity-vs-Alignment}, the optimum value of $k$ ranges from 15 to 25 for short reads when using Lighter. Performing an exhaustive search that requires alignment algorithms and entire reads is too time consuming. The optimum parameter configuration being dataset- and tool-dependent, further aggravates this process. For example, scanning over all $k$-mer values between 15-25 for dataset D6 and estimating the corresponding alignment rate with Bowtie2 took 1.5 hours on a server with 8 CPU cores and 12GB of RAM. In comparison, \name is 80x to 275x faster than Bowtie2 and was able to scan the same range in less than a minute. 

\subsection*{State-of-the-art and our contribution}
State-of-the-art automated EC tool tuner \textit{Athena}~\cite{abdallah2019athena}
leverages NLP techniques to automatically compute the best value of $k$ for error correction. However, Athena suffered from 
poor performance on datasets with larger read lengths, starting to degrade beyond reads of length 200 (as shown in Table~\ref{tab:corr}). 
Further, as a result of the inability of RNNs to encode long-term dependencies, along with the character-level LMs being space inefficient, \textit{Athena} failed to tune EC tools on long-read datasets.
To address this problem, we engineered word-level transformer networks to deploy attention mechanisms, at different encoding granularities. This leads to improved alignment performance for lower read lengths and the ability to handle higher read lengths. 
With \name's character- and word-level encodings encapsulated in transformer networks, our hypothesis of leveraging more sophisticated architectures for noisier and longer reads was validated. \textit{Specifically, \name either derived the optimal $k$-value or values corresponding to alignment rates within 0.31\% of the optimal, plus improved runtimes of 18$\times$ faster than the best-in-class---Athena.}
Having improved the correlation rates for short reads, we moved on to solve the harder problem of finetuning EC tools for longer reads. This is because third-generation sequencing technologies have low signal-to-noise ratios and a high error rate in base calling, with rates as high as 15\% and 40\% for PacBio and Nanopore data, nearly two orders of magnitude higher than second-generation sequencing data.
Athena used a character-level RNN in its algorithmic suite. We find through detailed analysis that character-level models, in contrast to word-level models, provide weaker correlations between perplexity and alignment rate, making them unsuitable to use over longer and/or noisier datasets. Intuitively, with word-level LMs, we expand the vocabulary size ($4^W$) and allow the model to learn identifiable patterns in the data, where $W$ is the word length. 

Additionally, the much slower inference [18$\times$ slower] of Athena (see Table~\ref{runtime}) creates a bottleneck for downstream genomic analysis.
Furthermore, as indicated before, RNN-based language models have the fundamental flaw of being unable to capture long-term dependencies in sequences~\cite{trinh2018learning}; hence it was unable to autotune parameters for {\textit{long-read}} EC tools. 



\begin{figure}
\centering
\includegraphics[scale = 0.30]{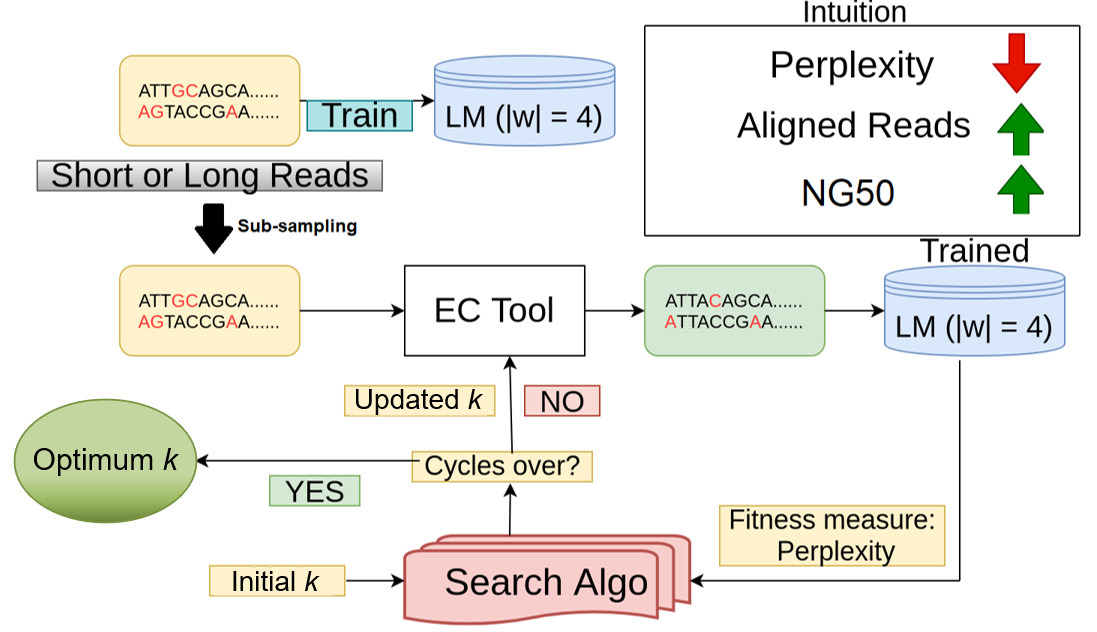}
\caption{Workflow of \name. \name's high level intuition is that by reducing the perplexity metric for the generated reads, we increase the alignment rate and the assembly quality (NG50). Our evaluation shows that a word-level Transformer LM, with a word length of 4 ($|w|$ = 4) works best across all datasets and tools. Our algorithmic suite works on both NGS short reads and PacBio and Nanopore long reads.}
\label{fig:overview}
\end{figure}

Our innovation, \name, schematically outlined in Figure~\ref{fig:overview}, employs a variant of Transformer networks~\cite{vaswani2017attention} to perform automatic tuning of EC tools on \textit{both} more error-prone, third-generation long reads and second-generation short reads. The original transformer used by 
Vaswani \textit{et al.} is causal and directional, 
\ie, the predictions for position $i$ can depend only on the elements at positions less than $i$. However, in our problem context of genomic error correction, such directionality is not apt. Therefore, we modify the encoder-decoder architecture in the transformer LM and allow it to train in a non-directional sense.

In summary, \name makes the following contributions:
\begin{enumerate}
    \item We propose a mechanism to automatically select the best configurations for $k$-mer based error correction tools. Our mechanism handles both short and long reads by modifying Transformer Language Models~\cite{vaswani2017attention} by disabling the masks to use non-directional word-level training, improving $k$-mer selection accuracy by 35\% over unidirectional character-level RNNs used by prior works. Further, these non-directional models were found to be equivalent to the bi-directional models in our experiments and thus we disabled the masks, reducing the number of model hyperparameters.
    
    
    
     \item We propose several runtime improvements such as leveraging parallelizable computations in self-attention layers, and utilizing JIT (Just-in-Time) compilers in order to make better optimizations based on the underlying GPU hardware. These improvements lead to efficient utilization of resources and 18$\times$ lower inference time than the state-of-the-art tool Athena for choosing the best configuration parameters for genomic error correction (see Table~\ref{runtime}).

    \item We evaluate the generalizability of \name on second- and third-generation sequencing technologies, specifically, Illumina short reads, and PacBio and Nanopore long reads, validating that \name can optimize both short and longer, noisier reads. 
\end{enumerate}
Furthermore, \name reports a negative correlation in each and every dataset (unlike Athena, which has a positive correlation of +0.95 on D6) with an average of -0.92 over D1 to D7, degrading with longer read lengths.
Our experiments have shown that the highest negative correlation is observed when word-level Transformer LMs are used, rather than character-level.
Intuitively, with word-level LMs, we expand the vocabulary size 
and allow the model to learn identifiable patterns in the data.



\subsection*{Language Modeling}
Language modeling has played an integral role in improving speech recognition, text analysis, and sentiment analysis. A Language Model (LM) is a probability distribution over a sequence of tokens (\eg words or characters), using which we can identify the sequences that are more likely to occur~\cite{zhai2008statistical, agarwal2011sentiment, elaraby2016deep, lhoussain2015adaptating}. 
LMs can be count-based (\eg estimating N-gram probabilities via counting and subsequent smoothing)~\cite{siivola2005growing} or continuous-space~\cite{schwenk2006continuous} language modeling, such as the use of Recurrent Neural Networks (RNNs). 
However, several papers have pointed out various drawbacks of using RNNs (see the next subsection). As an alternative, we make use of computationally efficient transformer LMs and further modify its architecture for speedup and parallelization. 

\subsection*{N-Gram-Based Language Modeling}
N-Gram models~\cite{brown1992class} are used to predict the probability of observing a token [\textit{$W_i$}] after the series of all already observed tokens [${W_0, \ldots W_{i-1}}$] (\ie, context). 
This is done for all possibilities present in the vocabulary. However, being a computationally expensive task, the above probability is approximated using Equation~\ref{ngram_calc}.
\begin{equation}
\footnotesize
  \begin{aligned}  
    P(W_0, W_1, ..., W_{m}) &= \prod_{i=1}^{m}  P(W_{i} |  W_{i-1}, ..., W_{1}) \\
  &\approx \prod_{i=1}^{m}  P(W_{i} |  W_{i-1}, ..., W_{i-n})
  \end{aligned}
  \label{ngram_calc}
\end{equation}

\noindent where \textit{n} represents the context. As we increase $n$, the accuracy increases, albeit at the cost of increasing complexity and training time. The conditional probabilities are stored as a lookup table, leading to a higher memory complexity. The size of the model on $D7$ for the N-gram LM is 2 GB while the size of the RNN model is only 3.5 MB for $D7$ (a description of the datasets is given in Table~\ref{tbl:Datasets_Description}). Moreover, N-gram-based models suffer from increased data sparsity when trained on longer contexts. As N increases, the number of possible word combinations becomes $2^N$ and the amount of data becomes inadequate to train the N-grams. In contrast, Neural LMs (such as RNNs) overcome these limitations and are preferred over N-gram LMs~\cite{kombrink2011recurrent}. 

\subsection*{RNN-Based Language Modeling}
\label{rnn}RNNs are a class of LMs that rely on an internal ``state'' to make predictions, usually consisting of three or more layers. 
The input layer is given an input vector $ x_{t} $. This may be a one-hot encoding vector, such as in our case, of the $ t^{th} $ word or character. The hidden layer represents the memory of the network: the internal ``state'' $ s_{t} $. A non-linear function \textit{f} (\eg tanh) is applied to the previous hidden states $ s_{t-1} $ and $ x_{t} $ (see Eq \ref{rnn_calc})
\begin{equation}\label{rnn_calc}
\centering{
s_{t} = f(U x_{t} + W s_{t-1}).
}
\end{equation}
Where $U$ and $W$ are the input and state weights respectively.
However, RNNs suffer from the problems of vanishing and exploding gradients~\cite{pascanu2013difficulty}. Due to the sequential nature of gradient computation results, training of RNNs is computationally demanding. Despite recent advances in training RNNs, \textit{capturing long-term dependencies in sequences remains a fundamental challenge}~\cite{trinh2018learning}. As the sequence length increases, the path length of information flow increases, resulting in greater likelihood of the information being corrupted. The inherent sequential nature of RNNs precludes parallelization within training examples, which becomes critical at larger sequence lengths, as memory constraints limit batching across examples~\cite{vaswani2017attention}. LSTMs~\cite{hochreiter1997long} and GRUs~\cite{chung2014empirical} are alternatives but the \textit{computation cannot be parallelized} because of the sequential nature of operations in that the internal state at any time step $t$ depends on the state of previous time step $t-1$. 
This remains a fundamental issue in all sequential ML models. Various architectures have been proposed to solve these problems. 
The common target of all of them is to use attention mechanisms to model probability distributions~\cite{vaswani2017attention,liu2019roberta}.
 
\subsection*{Transformer-Based Language Modeling}
 Attention mechanisms are a way of selectively focusing on certain parts of a sequence~\cite{luong2015effective}. Specifically, \textit{self attention} relates different positions of a single sequence in order to compute a representation of the sequence and has been used in a variety of tasks~\cite{cheng2016long}, \eg abstractive summarization~\cite{parikh2016decomposable}, textual entailment~\cite{paulus2017deep}, and learning task-independent sentence representations~\cite{lin2017structured}. We incorporate self-attention in the Transformer network architecture. Learning long-range dependencies is a key challenge in many sequence transduction tasks. One key factor affecting the ability to learn such dependencies is the length of the paths that forward and backward signals have to traverse in the network. The shorter these paths between any combination of positions in the input and output sequences, the easier it is to learn long-range dependencies. In the transformer, this value is constant $\mathcal{O}(1)$. Furthermore, because the transformer contains no recurrence and convolutional operations, in order for the model to make use of the ordering of the input sequence, positional encoding of the input is performed. This encoding is done by using \textit{sine} and \textit{cosine} operations of different frequencies as follows:
 \begin{equation}
      PE_{(pos,2i)}=\sin(pos/10000^{2i/d_{model}})
 \end{equation}
 \begin{equation}
    PE_{(pos,2i+1)}=\cos(pos/10000^{2i/d_{model}})
 \end{equation}
 Here, $d_{model}$ represents the embedding dimension, $pos$ is the position, and $i$ is the dimension index. Recurrence relations generate states $s_{t}$ that depend on the previous state $s_{t-1}$ and the input at time $t$. This sequential nature of computations precludes parallelization, hence the removal of recurrence relations from the Transformer allows it to be parallelized~\cite{vaswani2017attention}. 
\subsection*{Genomics Preliminaries}
This section gives brief definitions for a few genomics concepts that we focus on throughout the paper, such as error correction, alignment, and assembly. \texttt{Error correction} represents the process of identifying erroneous nucleotide base pairs in genomics reads and converting them to the correct pairs. This process is applied as a pre-processing step for the sequenced reads before further analysis is performed. It was shown that error correction can significantly improve the performance for all tasks downstream the analysis pipeline. \texttt{Alignment} is the process of matching the sequenced reads to a continuous error-free reference genome and identifying which segment in the reference genome matches a given read the best. \texttt{Assembly} is the process of merging the sequenced reads together into longer sequences in order to retrieve the original sequence. The design of \name focuses on improving the performance of error correction algorithms by identifying the best configuration parameters, which significantly improves the alignment rate and the assembly quality.

\subsection*{Perplexity of the Language Model}
\label{sec:ppl}

Perplexity is a metric that measures 
the goodness-of-fit of a sequence given an LM. The LM represents the probability distribution over the entire dataset~\cite{azzopardi2003investigating}. Perplexity can be thought of as the probability of selecting a given word uniformly from the effective vocabulary, given some context. Thus, a lower perplexity indicates that the LM is better at making predictions. If an LM learns a uniform probability distribution over a given piece of text, the perplexity generated would be equal to the actual vocabulary size. The smaller the perplexity score is, compared to the actual vocabulary size, the better is the LM at learning patterns that occur with a high probability in the given text. In our context of genome error correction, the error correcting tools are expected to convert most of the insolid $k$-mers to solid ones. Since solid $k$-mers have a high frequency of occurrence, more the conversion of insolid to solid $k$-mers takes place, lower is the effective vocabulary size of the corrected data. This can be quantified using the perplexity metric.
Mathematically, 
perplexity of a sentence (Eq~\ref{eq:4}) is the inverse probability of the test set, normalized by the number of words~\cite{azzopardi2003investigating}. 
 \begin{equation}
\small
\label{eq:4}
\begin{aligned}
 \small
  PP(W) & 
  \approx \sqrt[m]{\frac{1}{\prod_{i=1}^{m}P(W_{i} | W_{i-1},...,W_{i-n})}}
\end{aligned}
\end{equation}
It is clear from \eqref{eq:4} that minimizing perplexity maximizes the probability of the observed set of $m$ words from $W_{1}$ to $W_{m}$.
\rev{Various models estimate the perplexity metric differently. However, they achieve the same purpose, which is estimating the correctness of a sequence given the trained probability distribution. Although the perplexity for Transformers and N-Grams is calculated in different ways, it gives us a measure of probability of a sequence given a trained LM. The Kullback-Leibler (KL) divergence (Eq~\ref{eq:kl}) is a fundamental equation of information theory that quantifies the proximity of two probability distributions. It specifies in bits how close a probability distribution $P_i$ is to another distribution $Q_i$~\cite{shlens2014notes}.
\begin{equation}
\label{eq:kl}
D \left( P \middle\| Q \right) = \sum\nolimits_{i} P_i\log_2 \frac{P_i}{Q_i}
\end{equation}
A metric similar to the KL divergence between two probability distributions is the cross entropy (Eq~\ref{eq:ce}). It is also used to measure the similarity between two probability distributions. However, there are subtle differences between the metrics. Cross entropy gives the average code length needed to represent one distribution by another distribution, and the excess code needed over the optimal coding is given by the KL divergence~\cite{sbert2018some}.
\begin{equation}
\label{eq:ce}
H(P,Q) = -\sum\nolimits_{i} P_i\log_2 Q_i = H(P) + D \left( P \middle\| Q \right)
\end{equation}
}
In \name, the Transformer LM uses the perplexity metric (Eq~\ref{eq:5}), which is derived to be the exponential of the cross-entropy loss, where $\hat{y}$ is the predicted next character---the output of the LM---and $|V|$ is the vocabulary size used during training. 
 Therefore, by using perplexity, we incorporate information-theoretic principles into our LM, and in effect, ensure that by minimizing the perplexity, we maximize the similarity between the ground truth and the predictions made by the LM.
\begin{equation}
\label{eq:5}
Perplexity = e^{CE(y, \hat{y})} = \exp{(- \sum_{i=1}^{|V|} p(y_{i})\log(p(\hat{y_i})))}
\end{equation}

Note that training our LM model on the uncorrected reads can still capture an accurate probability distribution, which we use to measure the quality of the corrected reads. The reason is that genomic reads usually have a high coverage (typically 30X or more), which makes the number of occurrences for an erroneous nucleotide much lower than that of the correct nucleotide. This means that the correct genomic reads still dominate the distribution.


Now, we discuss the key design elements of \name and the corresponding challenges addressed by each element.
\subsection*{Transformer-Based Language Modeling in \name}

 The attention function (Eq~\ref{eq:at}) used in a Transformer can be described as mapping a query and a set of key-value pairs to an output, where the query, keys, values, and output are all vectors. The output is computed as a weighted sum of the values, where the weight assigned to each value is computed by a compatibility function of the query with the corresponding key. Specifically, scaled-dot product attention is done by taking dot product of \textit{Query} $Q$ and \textit{Key} $K$ vectors. This is then divided by $\sqrt{d_k}$, where $d_k$ is the dimension of $K$. This scalar value is multiplied by value vector $V$, and a softmax is taken to get a probability distribution.
 \begin{equation}
 \small
 Attention(\textbf{Q},\textbf{K},\textbf{V}) = \text{softmax}((\frac{\textbf{QK}^{T}}{\sqrt{d_k}})\textbf{V})
 \label{eq:at}
 \end{equation}
 
 Using multiple heads, rather than one, allows the language model to attend to information from different representation subspaces rather than from a single one. In the Transformer, the Attention module repeats its computations multiple times in parallel. Each of these is called an attention head.
 Because the layer after the concatenation is a feed forward network with an input dimension that does not depend on the number of heads, we take a linear combination of the concatenated vectors by multiplying with a matrix $W^O$ so that the resulting vector has the same dimension as the input dimension of the feed forward network.
\begin{equation}
\small
 MultiHead(\textbf{Q},\textbf{K},\textbf{V}) = \text{Concat}(head_1,\ldots ,head_h) . \textbf{W}^{O}
\end{equation}
where $head_i$ = $Attention(\textbf{QW}_i^{Q},\textbf{KW}_i^{K},\textbf{VW}_i^{V})$.
The architecture described by~\cite{vaswani2017attention} introduces masks in the self-attention layers. 
The self-attention sublayer in the decoder stack has been designed to prevent positions from influencing subsequent positions. This masking ensures that the predictions for position $i$ can depend only on the known outputs at positions $<$ $i$.
\textit{We modify this in \name to remove these masks, allowing the representations to encode both past {\em and} subsequent nucleotides. This is important because each nucleotide in a $k$-mer provides critical information, so the encoding of a given nucleotide in $k$-mer should have information about what is to the left and to the right of it.}
BERT~\cite{devlin2018bert} successfully  alleviates the  unidirectionality constraint of the Transformer, by using deep bidirectional transformers. Drawing inspiration from BERT, we considered implementing two parallel architectures with masks activated---one for the forward direction and the other for the reverse direction. This would allow the implementation of bi-directionality into the LM. \textit{However, empirically, we have shown that disabling masks and bi-directionality give the same results. Hence, we simply disable masks in order to reduce the number of parameters in the model.}
RNNs compute an internal state at each time step $s_t$, which depends on the previous state $s_{t-1}$ and the current input at time $t$. The Transformer, which gets rid of recurrent layers, are not restricted by this recursive relation, which must be computed one after the other, allowing parallelization.
The use of parallel attention modules has previously been used in reducing training time and establishing a new state-of-the-art model for English to German translation~\cite{medina2018parallel}. The ability of the Transformer networks to be parallelized opens up greater scope for optimizations in the machine-translated code.

\subsection*{Language modeling of genome sequences}
In general, LMs are trained to statistically determine if a sequence of tokens (\eg words in the NLP domain, or base-pairs in our usecase) contains erroneous (insolid or untrusted) tokens. We make use of the fact that sequencing technologies require sequencing every base-pair multiple times to provide a better and higher coverage of the genome. Accordingly, the \rev{number of erroneous reads becomes far less than the number of error-free reads}, allowing correct $k$-mers to outweigh erroneous ones. We train an LM on the set of reads before correction, expecting the correct sequences to mask the effect of the erroneous ones during the training phase. After correction of the reads by an EC tool, the resulting reads are tested against the trained language model. The evaluation metric chosen is \textit{perplexity} that is used to model the \textit{goodness of fit} of the corrected reads compared to the learned LM. The lower the perplexity, the better the fitness of the reads to the probability distribution captured by the LM. 

Following this argument, the error-free reads with plenty of solid $k$-mers are expected to produce a lower perplexity than erroneous reads with insolid $k$-mers, and this is exactly what we observed in our experiments as seen in Table~\ref{tab:err_ppl}. We generated a dataset with the NanoSim~\cite{yang2010reptile} simulator using the reference genome of \textit{E. coli}, strain K-12, substrain MG1655, with the characteristics of Nanopore sequencing~\cite{branton2010potential}. The generated data set had 588 reads labeled ``unaligned'' (\ie with nucleotide-level error rates over 90\%) out of 10,000 total reads.
We tested a word-level transformer model with word length = 4, on the entire dataset and separately on the erroneous and correct reads. From Table~\ref{tab:err_ppl}, it is clear that correct reads generate a lower perplexity than the erroneous ones for multiple values of $k$.
We use this idea to optimize the performance of EC tools by varying the input parameter $k$, and select the $k$-value that minimizes the perplexity. We have validated the results by performing alignment with the reference genome for the erroneous \textit{vs}. corrected reads. The corrected reads showed a higher alignment rate indicating a more effective correction of the sequences. 

\begin{table}[]
    \centering
    \caption{Effect of erroneous sequences on perplexity for multiple $k$-values on reads: These reads are generated by NanoSim using the $E. coli$ reference genome. $PPL_{err}$, $PPL_{corr}$, and $PPL_{total}$ denote the perplexity scores on erroneous and error-free reads, and the entire dataset (\ie erroneous and error-free sequences.)}    
    \begin{tabular}{|c|c|c|c|}
        \hline
        $k$ & $PPL_{err}$ & $PPL_{corr}$ & $PPL_{total}$ \\
        \hline
        15 & 1073.6 & 943.5 & 952.9 \\
        \hline
        37 & 1072.8 & 944.5 & 946.8 \\
        \hline
        81 & 1072.8 & 973.4 & 956.3 \\
        \hline
    \end{tabular}
    \label{tab:err_ppl}
\end{table}

\noindent {\bf Challenges in adapting LM for genomic reads}. There are a few differences in the training of an LM on the English language (or on any other language for that matter) and a genome sequence. \textbf{First}, a genome sequence has no concrete concept of words unlike an actual language. One may train a character-level model on it, but the training may be negatively affected by the small vocabulary size of sequences consisting of only A, T, G, and C. Also, one may use a fixed word length and pre-process the sequences to generate words, by using a suitable stride value. We have reported the effects of word length on training in our results that a word length of 4 leads to stronger negative correlation between perplexity and alignment rate, in contrast to smaller or larger word lengths as too small or too large vocabulary sizes can make learning patterns difficult for the LM.
\textbf{Second}, a genome sequence has no concept of punctuation marks such as periods. The sequenced reads, although having a finite read length, do not necessarily follow in succession. In the English language, the word following a period proceeds the word preceding it. In contrast, in a genome sequence, successive reads may or may not follow each other. Therefore, the batch creation for training or testing cannot be done arbitrarily. It has to be ensured that successive reads do not affect the training of other reads and that there are no unjustifiable breaks within a read during batch creation. \rev{Specifically, every read is tokenized into chunks of 100 base-pair length, before being fed to our transformer-based model. To preserve the overlap between the produced chunks, we use a sliding window with a step size (\textit{a.k.a.} stride length) of 50. If length of a read in not divisible by 100, the residual portion of the read is trimmed as they account for a small fraction of the read.} \textbf{Third}, genome sequences can be considered non-directional. A portion of a genome can as much be affected by the portion of the genome preceding it as by the portion of the genome in succession. This is also true for the English language \cite{devlin2018bert} but most of the models available are directional. Recent developments have resulted in the transformer network architecture that we orchestrate in \name to conform to the non-directionality attribute of genomics reads.
\textbf{Fourth}, since we train our model on the data \textit{before} correction, expecting the solid $k$-mers to influence training more than the erroneous ones (by outnumbering the erroneous ones),
it has to be taken care that the batch size during training $\geq$ the batch size during testing.
A larger batch size during training will help circumvent the effect of noisy or erroneous data, and help the model better learn the patterns occurring with high frequency. On the contrary, a smaller batch size during the testing phase will help identify sequences that have not occurred frequently during the training. Following the above rationale, we justify the use of NLP techniques to model genome sequences for evaluating the performance of EC tools.

\subsection*{Search Heuristic}
\label{sec:searching}
Our aim is to find the $k$-value that will minimize the perplexity of the corrected sequence. This ensures that the sequence has the highest chance of occurring, signaling that the EC tool is working optimally. The perplexity function calculated using the LM is denoted by $f$ (Eq~\ref{eq:perplexity_fun_hill_climb}). Although we experiment with $k$-mer-based approaches, any configuration parameter can be used in the search depending on the tool used, such as the maximum number of passes per read in Jabba~\cite{miclotte2016jabba}, or the value of the maximum boundary difference to split the subcontigs in HALC~\cite{bao2017halc}. Whatever the configuration parameter may be, our search heuristic uses perplexity as a measure of ``fitness''. 
Our search heuristic aims to find the configuration parameter with the minimum perplexity, and in doing so, maximizes the alignment rate and the assembly quality.\\
\rev{
\begin{equation}
\label{eq:perplexity_fun_hill_climb}
\begin{aligned}
  k_{opt}  = arg\min_{k_i}  Perplexity = arg\min_{k_i} f(LM, D_0, k_i)
\end{aligned}
\end{equation}}
Here, LM: trained language model, $D_0$: uncorrected read set, and $k$: the configuration parameter we wish to tune.

In \name, we apply a Simulated Annealing (SA)-based search that improves an initial solution by walking randomly in the space of possible solutions and gradually adjusts a parameter called ``temperature''. At high temperature, the random walk is almost unbiased and it converges to essentially the uniform distribution over the whole space of solutions. As the temperature drops, each step of the random walk is less likely to explore sub-optimal choices. 
Unlike a Hill Climbing-based approach used by prior work Athena, Simulated Annealing does not get stuck at local optimum, and converges faster to the global optimum~\cite{kalai2006simulated}.
The problem with local optima is partially mitigated in Athena by multiple runs of the search with various initializations but this comes at the cost of higher execution times.
\section*{Results}
\revv{Theoretically, \name can help tune any EC tool, but for the sake of evaluation, we have used Lighter for short reads correction and LoRDEC for long reads correction.}
\subsection*{Short Reads}
\label{sec:short_reads_results}
We begin with the demonstration of the effectiveness of \name on short reads datasets. The details of the seven NGS datasets used are given in Table \ref{tbl:Datasets_Description}. These are Illumina short reads and have been used in multiple prior studies (\eg~\cite{yang2010reptile,song2014lighter}) with reference genomes available (used to evaluate the quality of error correction). The datasets have different read lengths (from 36bp to 250bp) and different error rates \revv{($<$ 3\%)} \revvremoved{(from $<$ 3\% to 43\%)}.  
After correction using EC tool Lighter~\cite{song2014lighter}, we run the Bowtie2 aligner~\cite{langmead2012fast} and measure the Alignment Rate. A higher value implies superior error correction. We do an exhaustive search over various $k$-values to see whether \name produces optimal values. Table \ref{tb1:Lighter-Blue-Racer-Perplexity-vs-Alignment} summarizes our findings. We see that the Transformer word-level LM performs slightly better than the character-level variant; the former predicts the correct value in 5 datasets while the latter makes accurate predictions in 4 datasets; these out of a total of 7 datasets. Furthermore, when optimal parameters are not chosen, the alignment rate lies within 0.31\% of the best possible alignment rate for both char- and word-level models. 

\begin{table}[H]
\begin{center}
\caption {NGS short-read datasets' description with coverage (estimated per Illumina's documentation), number of reads, read lengths, genome type, and the Accession \#.} 
\label{tbl:Datasets_Description}
 \scalebox{0.85}{
 \begin{tabular}{|c|c|c|c|c|c|c|} 
 \hline
 Dataset &  Coverage & \#Reads & Read Length & Genome Type &  Accession Number \\ [0.5ex] 
 \hline
 \multirowcell{1}{D1} & 80$\times$ & 20.8M &  \multirowcell{1}{36 bp} & \multirowcell{1}{\textit{E. coli} str. K-12 substr} & SRR001665 \\
 \hline
 D2 & 71$\times$ & 7.1M & 47 bp & \multirowcell{1}{\textit{E. coli} str. K-12 substr} &  SRR022918  \\ 
 \hline
 D3 & 173$\times$ & 18.1M & 36 bp & \multirowcell{1}{\textit{Acinetobacter} sp. ADP1} &  SRR006332   \\ 
  \hline
  D4 & 62$\times$ & 3.5M & 75 bp & \multirowcell{1}{\textit{B. subtilis}} &  DRR000852  \\
   \hline
  D5 & 166$\times$ & 7.1M & 100 bp & \multirowcell{1}{\textit{L. interrogans C} sp. ADP1} &  SRR397962    \\
 \hline
   D6 & 70$\times$ & 33.6M & 250 bp & \multirowcell{1}{\textit{A. thaliana} } &  ERR2173372    \\
    \hline
   D7 & 67$\times$ & 202M & 101 bp & \multirowcell{1}{\textit{Homo sapiens} } &  SRR1658570    \\
 \hline
\end{tabular}}
\end{center}
\end{table}

A comparison of \name and Athena can be seen in Table \ref{tab:corr}. We notice that \name's Transformer-based model produces a higher correlation with the alignment rate compared to the RNN-based model used in  Athena for longer reads (\ie, \textit{D6} with 250 bp). This shows that more sophisticated NLP architectures are better suited for longer and noisier reads as is typical of the long-read sequences produced by PacBio and Nanopore sequencers. Recall that our refined NLP architecture 
uses parallelizable attention mechanisms in place of sequential recurrent layers, which affords higher optimizations. Furthermore, by restricting the maximum path length to $\mathcal{O}(1)$, we have prevented the corruption of information. This is unlike RNNs, in which the path length depends on the length of the input sequence.

\begin{table}[h!]
\centering
 \caption {\name evaluated on 7 Illumina (Table \ref{tbl:Datasets_Description}) short read datasets. We test both the Transformer character- and word-level LMs. We observed that the best performance was attained by using the Transformer word-level LM with word length 4 ($|w|$=4). For the Transformer word-level LM ($|w|$ = 4), \name finds either the optimal value or a value with an alignment rate within 0.31\% of the theoretical best, consistent with the reported results by Lighter (Figure 5 in~\cite{song2014lighter}). These slightly sub-optimal configurations are an artefact of sub-sampling.} 
\scalebox{0.62}{
\small
\begin{tabular}{ |c|c|c|c|c|c|c|c|c|c| } 
\hline

{\textbf{Dataset}} & {\textbf{Read Length}} & \multicolumn{2}{c|}{\makecell{\textbf{Exhaustive Search}}} & \multicolumn{2}{c|}{\textbf{ \name (Char) }} &\multicolumn{2}{c|}{\textbf{ \name (Word) }} & \multicolumn{2}{c|}{\textbf{  Athena (RNN) }}\\ 
 \hline
 					 & \makecell{\textbf{Base Pairs}\\
 					 \textbf{$bp$}} & \makecell{\textbf{Selected}\\ \textbf{$k$}} & \makecell{\textbf{Alignment}\\ \textbf{Rate (\%)}}& \makecell{\textbf{Selected} \\ \textbf{$k$}} & \makecell{\textbf{Alignment}\\ \textbf{Rate (\%)}} & \makecell{\textbf{Selected} \\ \textbf{$k$}} & \makecell{\textbf{Alignment}\\ \textbf{Rate (\%)}} & \makecell{\textbf{Selected} \\ \textbf{$k$}} & \makecell{\textbf{Alignment}\\ \textbf{Rate (\%)}} 
\\ \hline
 					 \textbf{D1} & \textbf{36} & \textbf{17} & \textbf{98.95}\% &  \multicolumn{2}{c|}{Same as Exhaustive Search} & \multicolumn{2}{c|}{Same as Exhaustive Search} &
 					 \multicolumn{2}{c|}{Same as Exhaustive Search} 
\\ \hline
     					  \textbf{D2} & \textbf{47} & \textbf{15} & \textbf{61.42}\% &  \multicolumn{2}{c|}{Same as Exhaustive Search} & \textbf{19} & \textbf{61.27}\% & \textbf{k=17} & \textbf{61.15}\% 
\\ \hline
     				 \textbf{D3} & \textbf{36} & \textbf{15} & \textbf{80.44\%} &   \textbf{k=17} &  \textbf{80.39\%} & \multicolumn{2}{c|}{Same as Exhaustive Search} &   \textbf{k=17} &  \textbf{80.39\%}  
\\ \hline                        
                        \textbf{D4}& \textbf{75} & \textbf{17} & \textbf{93.95\%} &   \textbf{k=15} &  \textbf{93.72\%} & \multicolumn{2}{c|}{Same as Exhaustive Search} & \multicolumn{2}{c|}{Same as Exhaustive Search} 
\\ \hline                           
                         \textbf{D5}& \textbf{100} & \textbf{17} & \textbf{92.15}\% &  \multicolumn{2}{c|}{Same as Exhaustive Search} &  \multicolumn{2}{c|}{Same as Exhaustive Search} & \textbf{k=25} & \textbf{92.09}\%  
\\ \hline                           
                         \textbf{D6}& \textbf{250} & \textbf{25} & \textbf{86.16}\% &  \multicolumn{2}{c|}{Same as Exhaustive Search} &  \multicolumn{2}{c|}{Same as Exhaustive Search} & \textbf{k=17} & \textbf{85.63}\%  
\\ \hline                           
                         \textbf{D7}& \textbf{101} & \textbf{17} & \textbf{40.53\%} &   \textbf{k=15} &  \textbf{40.24\%} &   \textbf{k=15} &  \textbf{40.24\%} &  \multicolumn{2}{c|}{Same as Exhaustive Search}\\ 
  \hline
  
  \end{tabular}}
\label{tb1:Lighter-Blue-Racer-Perplexity-vs-Alignment}
\end{table}

Next we measure the improvement of \name's pipeline on the assembly quality. In short, assembly is the process of merging the reads together to make longer reads (\textit{a.k.a} contigs), and the longer the constructed contigs, the closer they are to the original sequence (consensus regions). We use SPAdes assembler~\cite{prjibelski2020using} to perform \textit{de novo} assembly for the reads before and after correction and then use QUAST~\cite{gurevich2013quast} on the generated contigs to estimate the assembly quality metric NG50, which we show in Table \ref{tab:corr}. \rev{The N50 is a metric that indicates that contigs equal to or larger than its value represent 50\% of the total assembly size, and the NG50 is similar to N50, but based on the estimated genome size.} NG50 is similar to mean or median of the generated read contigs but with a higher weight assigned to longer sequences, hence the higher the better. \rev{We notice that \name is able to improve the NG50  by a geometric mean of 5.08$\times$ across the 7 real datasets as opposed to Athena, which shows an improvement of 4.72$\times$. This shows the importance of using \name's pipeline to improve the performance for alignment and assembly tasks significantly.}

\begin{table}[h!]
    \centering
    \caption{\rev{A comparison between Athena and \name on short reads. The dataset is described in Table \ref{tbl:Datasets_Description}. We show the correlation between the perplexity metric and the alignment rate of the data after correction for \name vis-\`a-vis its closest competitor, Athena. On dataset \textit{D6} (greatest read length of 250 bp), Athena fails and has a positive correlation of +0.95 (instead of having a negative correlation). This is in line with the fact that RNNs are unable to model longer sequences. We also show the improvement of the assembly quality (in NG50) after tuning the EC tool with \name \textit{vs}. using the uncorrected reads.}}    
    \begin{tabular}{|c|c|c|c|c|c|c|c|}
        \hline
        Dataset &\makecell{  Read \\ Length}&\makecell{  Coverage} & \makecell{ Correlation \\ Athena } &  \makecell{ Correlation \\ \name } & \makecell{ NG50 \\ without \\ Correction } & \makecell{ NG50 \\ with \\ \name } & \makecell{ NG50 \\ with \\ Athena }\\
        \hline       
        D1 & 36 bp & 80$\times$ & -0.93 & -0.94 & 3019 & 6827 & 6827 \\
        \hline
        D2 & 47 bp & 71$\times$ & -0.97& -0.96 & 47 & \textbf{2254} & 2164\\
        \hline
        D3 & 36 bp & 173$\times$ & -0.92& -0.93 & 1042 & \textbf{4873} & 4164\\
        \hline
        D4 & 75 bp & 62$\times$ & -0.86& -0.97 & 118 & 858 & 858\\
        \hline
        D5 & 100 bp & 166$\times$ & -0.96& -0.98 & 186 & \textbf{3524} & 2799\\
        \hline
        \textbf{D6} & \textbf{250 bp} & 70$\times$ & \textbf{+0.95}& \textbf{-0.84} & 1098 & \textbf{1344} & 1237\\
        \hline
        D7 & 101 bp & 67$\times$ & -0.72 & -0.82 & 723 & 739 & \textbf{754}\\
        \hline
    \end{tabular}
    \label{tab:corr}
\end{table}


\subsection*{Evaluation on long reads}
\label{sec:long_reads_results}
For long-reads evaluation, we use the PacBio simulator~\cite{ono2013pbsim} to generate 10,000 reads with read lengths that vary between 2000 to 3000 base pairs and an accuracy of 78\%, which matches the characteristics of real continuous long reads (CLR). Due to the varying read lengths in the dataset, an additional pre-processing step is performed before training our LM. Specifically, every read is tokenized into chunks of length 100 base-pairs before being fed to our transformer-based model. To preserve the overlap between the produced chunks, we use a sliding window with a step size (a.k.a stride length) of 50. If length of a read in not divisible by 100, the residual portion of the read is trimmed as they account for a small fraction of the read. This fraction is upper bounded at 1.6\% in our evaluation.  


We use the reference genome of \textit{E. coli}, strain K-12, substrain MG1655. We used LoRDEC~\cite{salmela2014lordec} for error correction since it has proven to work well on PacBio reads and is computationally efficient, yielding results within just a few minutes. \rev{Since it is a hybrid error correction tool, the short reads we use are those of \textit{E. coli}, strain K-12, substrain MG1655, accession number SRX000429.} The LM was trained on the uncorrected reads, and tested on the corrected reads. The ground truth was generated by running alignment using Minimap2~\cite{li2018minimap2} and the statistics of the aligned reads were generated using paftools (provided in the Minimap2 repository). 
The experiments were repeated for a character-level transformer and a word-level transformer network, where the input sequences were converted into words during the pre-processing phase.

\begin{figure}
    \centering
    \includegraphics[scale = 0.26]{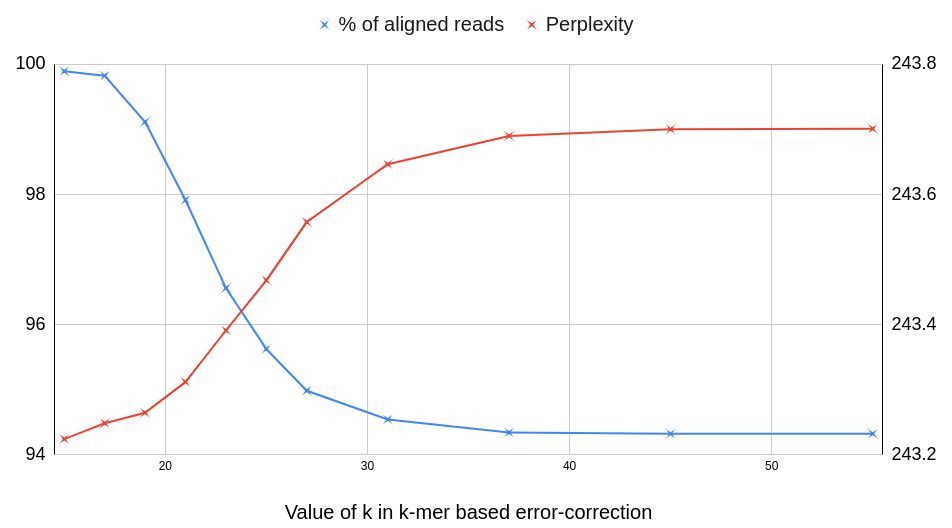}
    \caption{Variation of perplexity and \% of aligned reads with $k$ on simulated \textit{E. coli} PacBio reads, corrected by LoRDEC, mapped by Minimap2. The perplexity is computed by a Transformer word-level LM with $|w|=4$. A strong negative correlation between the two metrics is clearly observed. }
    \label{fig:ppl}
\end{figure}

\label{long}
Given that our data has 4 literals, A, T, G, and C, for a word length of $|w|$, the vocabulary size of the dataset is equal to $4^{|w|}$. The number of tokens in the model was set to $4^{|w|}$. The size of the word embedding vector was set to $4^{|w|}$ for $|w|<4$, and to 128 otherwise, because an embedding size larger than the number of tokens is not a sensible choice. The number of encoder layers was set to 4, same as that of the model trained on short reads. \rev{The hyperparameter \textit{bptt} represents back propagation through time, \ie the data is divided into chunks of size \textit{bptt}.} It was set to 99 after it was observed that it is not sensitive to the quality of training for a significant range, from 50 to 500. We use perplexity as our evaluation metric, and its correlation with the percentage of aligned reads after correction was computed. The results are reported in Table \ref{tab:word}. A more detailed result of variation of perplexity with \textit{k} is shown in Figure \ref{fig:ppl} for $|w|=4$. 

\begin{table}[]
    \centering
    \caption{Effect of word length on correlation between perplexity and the percentage of aligned reads for \textit{E. coli} PacBio simulated reads. Strong correlation is observed for higher $|w|$ values, with $|w| = 4$ producing the strongest correlation. The vocabulary size $4^{|w|}$ is represented by $|V|$.}
    \begin{tabular}{|c|c|c|c|c|c|}
        \hline
        $|w|$ & $|V|$ & Mean(PPL) & SD(PPL) & Correlation & Test time (s)\\
        \hline
        1 & 4 & 3.9997 & 0.00004 & -0.714 & 367.5\\
        \hline
        2 & 16 & 15.824 & 0.02298 & +0.825 & 990.1\\
        \hline
        3 & 64 & 62.176 & 0.06331 & +0.761 & 1439\\
        \hline
        \textbf{4} & \textbf{256} & \textbf{243.45} & \textbf{0.17869} & \textbf{-0.965} & \textbf{2026}\\
        \hline
        5 & 1024 & 951.70 & 2.86750 & -0.904 & 1862\\
        \hline
        6 & 4096 & 3724.7 & 21.5649 & -0.889 & 2472\\
        \hline
        7 & 16384 & 14675 & 133.207 & -0.882 & 2923\\
        \hline
        8 & 65536 & 59199 & 811.820 & -0.877 & 7150\\
        \hline
    \end{tabular}
    \label{tab:word}
\end{table}
In Table \ref{tab:word}, $|w|$ refers to the word length selected for training and $|V|$ is the vocabulary size of the data. The \textit{mean} and \textit{standard deviation} of the perplexity score has been calculated on the corrected data for \textit{$k$ = 15,17,19,21,23,25,27,31,37,45}, since beyond this value there was no change in the percentage of alignment for the corrected reads with the reference genome. Furthermore, a lower $k$-value is not usually recommended; in most cases, it is not supported by the error correction tool. The correlation between perplexity and the percentage of aligned reads was computed for the above mentioned values of \textit{k} after standard normalization of both, perplexity and the percentage of aligned reads. 
A negative correlation between perplexity and \% aligned reads was expected but it was interesting to see a positive correlation for $|w| = 2$ and $3$. A possible reason could be a low information gain in 2-mers and 3-mers; \ie a 2-mer may not be followed by a character with a high probability, and the prediction of a 3-mer from a given 2-mer appears to be random because of a small vocabulary size. We observe a low negative correlation for $|w|=1$. Although it is known that character-LMs perform well, the weak correlation is not good enough to eliminate the need for a reference genome. 
According to information theory, entropy is maximized when a uniform distribution is selected over a countable set~\cite{conrad2004probability}. The perplexity is the exponential of the cross entropy (Equation~\ref{eq:5}). Hence, the perplexity, in essence, models the effective vocabulary size, \ie, it equals the vocabulary size when the probability distribution of the tokens in a text data is uniform, and it is less than the vocabulary size when there are motifs present in the data, with identifiable patterns that occur with high probability.

\begin{table}[]
    \centering
    \caption{Results of \name evaluated on PacBio reads: $|w|$ denotes the word length used for training, $k$ is the selected $k$-value, $\textit{mapped reads}$ is the total number of reads out of the 10,000 reads that aligned with the reference genome after correction done by LoRDEC using the selected $k$-value.}
    \begin{tabular}{|c|c|c|}
        \hline
        $|w|$ & k & mapped reads \\
        \hline
        1 & 67 & 9432 \\ 
        \hline
        2 & 67 & 9432 \\ 
        \hline
        3 & 67 & 9432 \\ 
        \hline
        \textbf{4} & \textbf{15} & \textbf{9989 }\\ 
        \hline
        5 & 17 & 9982 \\ 
        \hline
        6 & 17 & 9982 \\ 
        \hline
        7 & 17 & 9982 \\ 
        \hline
        8 & 23 & 9656 \\ 
        \hline
    \end{tabular}
    \label{tab:simann}
\end{table}
From the results, we see that the difference between the vocabulary size and the mean perplexity grows with the word size. The standard deviation of perplexity also grows with word size. The task of our search heuristic (of finding the best $k$) is made easier since our LM can differentiate easily between perplexities corresponding to $k$-values. To eliminate the need for the ground truth, that is the reference genome, the sole requirement was a high correlation between perplexity and \% of aligned reads, and we see that strongest correlation was obtained for $|w| = 4$. Effectively, the word length only affects the effective \textit{bptt} value, given that a stride of 1 is chosen for converting text to words. It also helps set the vocabulary size, allowing a model to look for high probability motifs in different vocabulary spaces. For example, consider a sequence of length 10, AATCGGCGCT. Let us choose \textit{bptt=9}. A char-level model learns the following probability P(T$|$AATCGGCGCT) if trained on the given sequence. If we choose a word length of 4 with \textit{stride=1}, the data is pre-processed to generate the sequence, AATC ATCG TCGG CGGC GGCG GCGC CGCT. With \textit{bptt=6}, a word-level model learns the probability P(CGCT$|$AATC ATCG TCGG CGGC GGCG GCGC) that is the same as P(T$|$AATCGGCGC). Hence, a word-level model with \textit{word length=4, stride=1, bptt=6} effectively learns the same probabilities that a character-level model with \textit{bptt=9} learns, with the only difference being the vocabulary size. From our experiments, we have concluded that $|w|=4$ is the best choice for our use case. We believe further research can be done on how training LMs on genome sequences is affected by the word length. This will help us gain more insights on how to detect solid $k$-mers using NLP techniques. 
The results of our pipeline on PacBio simulated long reads, are reported in Table \ref{tab:simann}. We find that the results are consistent with our claims made above. For $|w| = 4$, our search heuristic selected the value of \textit{k} corresponding to the best alignment. 9989 out of 10000 reads aligned with the reference genome for \textit{k} = 15 which is selected only for $|w|=4$. For all other word lengths $|w|$, the optimal \textit{k} value for a minimum perplexity and the optimal \textit{k} value for maximum alignment rate are different, largely depending on the correlation between the two metrics.

Finally, we evaluate the improvement in assembly quality for the reads after correction using \name versus the original set of reads (uncorrected). We use SPAdes assembler to perform the assembly and generate the contigs, and we use QUAST measuring tool to estimate the NG50 for the generated contigs. The uncorrected reads had an NG50 of 26,034, whereas the corrected reads with LoRDEC+\name had an improved NG50 of 38,466, showing an improvement of 47\% (over uncorrected reads). This shows the importance of applying \name's pipeline to improve the performance for both alignment and assembly tasks. Compare this improvement to the improvement of 5.08$\times$ in assembly quality with short reads. This difference is due to the fact that assembly with shorter uncorrected reads is significantly worse than assembly with longer uncorrected reads. This is due to the fact that assembly is a harder problem with shorter reads, akin to piecing a jigsaw puzzle with a larger number of pieces. 


\revv{We evaluate \name on two real datasets --\textit{E. Coli K-12} PacBio reads (Accession number: SRX4909245) and \textit{Acinetobacter baumannii} Nanopore reads (Accession number: SRX11521539). The data was corrected with LoRDEC and assembly was done using Canu. 
The data was tokenized into words with $|w| = 4$ for the training and testing of the language model. Tables \ref{tab:real_pac} and \ref{tab:real_nano} show the evaluation results of \name on these datasets. We have performed experiments with the maximum range of $k$ values that each of LoRDEC and Canu could support on our machine.}

\revv{Since \name uses a heuristic-based optimization strategy (\ie Simulated annealing), finding the global optimum is not guaranteed. However, for all evaluated datasets, \name was able to tune different EC tools and find $k$-mer sizes that significantly improve the quality of the  corrected reads when the assembly results of the corrected reads are compared across the different $k$-mer sizes. For example, in Table \ref{tab:real_pac}, $k=17$ generates an NG50 value of 133690 which is significantly better than for example, $k=37$ which has an NG50 value of 32160.
Accordingly, it also improves the quality of assembly exemplified by the improvement in the NG50 metric. \name finds a value that is either identical or very close to the actual best $k$, but without relying on a reference genome.  For example, we see in Table \ref{tab:real_pac} that for the PacBio dataset, \name finds $k=17$ as the optimal $k$ whereas the actual best value was $k=19$. In Table \ref{tab:real_nano}, we see that for the Nanopore dataset, \name finds the $k$ value that results in the best assembly. Hence, we can conclude that \name successfully tunes LoRDEC on real long reads datasets as well.}

\revvremoved{Similar to the evaluation on PacBio simulated long reads, }We use Canu~\cite{koren2017canu} for error correction on Nanosim~\cite{yang2017nanosim} data. The Nanopore simulated reads are of \textit{E. coli}. Using \name, we try to find the best value of MhapMerSize, \ie the $k$-mer size for seeds in the MHAP part of Canu. Through our experiments, we found that a word length of 4 ($|w|=4$) gives the optimum value of $k=19$ and a corresponding 9,813 out of 10,000 reads, which aligned with the reference genome. 
This kind of configuration search is done in parallel using the Apache Spark framework as described in our prior work~\cite{ghoshal2015ensemble}. This is an example of what is called an embarrassingly parallel workload. 
The number of reads aligned and the corresponding value of MhapMerSize is given in Table~\ref{tab:canu}. The mean perplexity output from the LM is 293.11, with a standard deviation of 0.16836, and the correlation between the aligned reads and perplexity is -0.889. Hence, we show that \name works on both short- and long-read datasets.

\begin{table}[h!]
    \centering
    \caption{\rev{Results of \name evaluated on Nanopore simulated reads corrected using Canu. We use a word length of 4 ($w=4$) and find that the best value of the MhapMerSize comes out to be 19.}}
    \begin{tabular}{|c|c|}
        \hline
        k & mapped reads \\
        \hline
        11 & 9531 \\ 
        \hline
        13 & 9549 \\ 
        \hline
        15 & 9631 \\ 
        \hline
        17 & 9723\\ 
        \hline
        \textbf{19} & \textbf{9813} \\ 
        \hline
        21 & 9724 \\ 
        \hline
        23 & 9696 \\ 
        \hline
        25 & 9657 \\ 
        \hline
        27 & 9553 \\ 
        \hline
        29 & 9545 \\ 
        \hline
    \end{tabular}
    \label{tab:canu}
\end{table}

\begin{table}[] \caption{\revv{\name results on real PacBio $E. Coli$ $K$-12 reads. Simulated annealing finds $k=17$ as the best $k$ value which is also evident from the fact that it generated the minimum test perplexity, which is quite close to $k=19$ that generates the highest NG50 on assembly.}}
    \centering
    \begin{tabular}{|c|c|c|c|}
    \hline
        $k$ & Test PPL & NG50\\
        \hline
        15 & 240.57 & 101440\\
        \hline
        \textbf{17} & \textbf{240.53} & 133690\\
        \hline
        \textbf{19} & 240.59 & \textbf{181898}\\
        \hline
        21 & 240.62 & 166229\\
        \hline
        23 & 240.65 & 92537\\
        \hline
        25 & 240.67 & 92859\\
        \hline
        27 & 240.69 & 74776\\
        \hline
        31 & 240.70 & 58332\\
        \hline
        37 & 240.64 & 32160\\
        \hline
    \end{tabular}
   
    \label{tab:real_pac}
\end{table}

\begin{table}[]
    \caption{\revv{\name results on real Nanopore $Acinetobacter$ $baumannii$ reads. Simulated annealing here finds $k=13$ as the best $k$ value that also generates the highest NG50 on assembly.}}
    \centering
    \begin{tabular}{|c|c|c|}
    \hline
        $k$ & Test PPL & NG50\\
        \hline
        11 & 221.14 & 39916\\
        \hline
        \textbf{13} & \textbf{192.64} & \textbf{41358}\\
        \hline
        15 & 221.00 & 40271\\
        \hline
        17 & 222.93 & 40452\\
        \hline
        19 & 222.90 & 40478\\
        \hline
        21 & 222.76 & 40491\\
        \hline
        23 & 223.10 & 40484\\
        \hline
        25 & 222.96 & 40477\\
        \hline
        27 & 222.75 & 40489\\
        \hline
        31 & 222.69 & 40541\\
        \hline
        37 & 222.38 & 40541\\
        \hline
        45 & 223.01 & 40516\\
        \hline        
    \end{tabular}

    \label{tab:real_nano}
\end{table}

\subsection*{Just-in-Time compilation}
\begin{figure}[h] 
\includegraphics[scale=0.7]{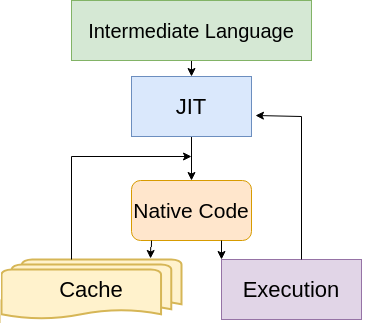}
\centering
  \caption{A JIT compiler, which leverages the qualities of both static compilers and interpreters, has information about the system during execution, enabling better optimizations.}
  \label{figure:JIT}
\end{figure}


We further improve the inference time of \name using Just-in-Time compilation~\cite{aycock2003brief}. The way bytecodes get converted to the appropriate native instructions for an application has a huge impact on the speed of an application. 
As seen in Figure \ref{figure:JIT}, they have access to run-time information such as input parameters, control flow, and target machine specifics. 
While JIT compilers provide good performance, their performance is often a blackbox~\cite{rompf2014surgical}. We employ JIT compilation in \name and find an immediate beneficial effect in the inference time. 
On using JIT compilers on Transformer LMs, we observe that the GPU utilization increases from 7\% to 50\% and power usage of the GPU increases by 22W. This increase signifies that the JIT compiler allows for data parallelization in the GPU; more \textit{data streams} and hence more GPU cores are being concurrently used. Clearly, by using JIT compilers, we have exploited the fact that the self-attention mechanisms can be parallelized, resulting in higher code runtime efficiency.

\begin{table}
\centering
 \caption {\name and Athena run on 7 Illumina short read datasets. The time for calculating perplexity has been reported, along with the read lengths and number of reads. We observe that on average our pipeline is 18$\times$ faster than Athena. This translates to 80$\times$ to 275$\times$ faster than estimating the alignment rate with Bowtie2.} 
\scalebox{0.75}{
\small
\begin{tabular}{ |c|c|c|c|c|c|c|c| } 
\hline

{\textbf{Dataset}} & {\textbf{Coverage}} & {\textbf{Genome}} & {\textbf{Read Length}} & {\textbf{\#Reads}} & {\textbf{Athena}} & {\textbf{\name}} & {\textbf{Speedup}}
\\ \hline
 					 D1 & 80$\times$ & \textit{E. coli} str. K-12 substr & 136bp & 20.8M & 98s &  5.5s & 17.8$\times$ 
\\ \hline
D2 & 71$\times$ & \textit{E. coli} str. K-12 substr & 47bp & 7.1M & 49s &  3s & 16.3$\times$ 
\\ \hline
 					      D3 & 173$\times$ & \textit{Acinetobacter} sp. ADP1 & 36bp & 18.1M & 69s &  4s & 17.25$\times$ 
\\ \hline                        
                        D4 & 62$\times$ & \textit{B. subtilis} & 75bp & 3.5M & 52s &  3s & 17.3$\times$ 
\\ \hline                           
                         D5 & 166$\times$ & \textit{L. interrogans C} sp. ADP1 & 100bp & 7.1M & 100s &  5.5s & 18.2$\times$ 
\\ \hline                           
                         D6 & 70$\times$ & \textit{A. thaliana} & 250bp & 33.6M & 400s &  23s & 17.4$\times$ 
\\ \hline                           
                         D7 & 67$\times$ & \textit{Homo sapiens} & 101bp & 202M & 960s &  63s & 15.2$\times$ \\
  \hline
  
  \end{tabular}}
\label{runtime}
\end{table}

While the GPU utilization for a transformer increases $\sim$7$\times$ after using a JIT compiler, for an RNN LM, the power usage and GPU utilization are not improved. The Transformer efficiently uses the GPU by allowing computations to occur in parallel. The consequences of using an LM, which has parallelizable attention mechanisms, along with optimizations made by the JIT compiler, is that inference time for the Transformer is improved. The inferences are made using an NVIDIA Tesla P100 16GB GPU. We find that \name (using Transformer) is 18$\times$ faster than Athena (using RNN) (Table~\ref{runtime}).  The gains are quite consistent across the different datasets that span a wide range of characteristics (such as coverage and number of reads). 

\section*{Discussion}
We have presented \name that can automatically tune EC tools by exploring their parameter search space, using data-driven techniques without the need for a reference genome for the sequence being corrected. We have made use of the fact that trained language models (LMs) generate lower perplexity when used to evaluate corrected reads. Further, given that finding the optimal configuration traverses a non-convex trajectory, we have used simulated annealing to reduce the chances of getting stuck in a local minima. Further, some EC tools may have performance-sensitive configuration parameters that may have interdependencies and, in such cases, a surrogate model that can encode dependencies, as in our recent work~\cite{mahgoub2017rafiki, mahgoub2020optimuscloud}, can be used. Finally, we have compared \name's findings with ground truth results on Illumina short reads and PacBio and Nanopore long reads. 
\name requires the user to first train an LM and then use the model to evaluate the corrected reads. Training can be a slow process and is directly proportional to the number of reads sampled for training. Our experiments have shown that \name works successfully even when as low as 5\% of the input data is used for training. The user, therefore, needs to decide between the sample size and the training time. This is because a larger sample size is expected to have a larger input coverage that can result in a more rigorously trained model. Training is however required to be done only once, and the trained model is used to evaluate the corrected reads multiple times as \name searches through the parameter space of the EC tool being used. Evaluation is fast, and works in the order of a few seconds, as reported in Table~\ref{runtime}. 

The EC tool needs to be run for every input parameter that \name suggests as it searches through the parameter space. The runtime of the EC tool depends not only on the tool and the dataset, but also on the value of the input parameter. For example, the runtime of LoRDEC increases exponentially as $k$ decreases. Therefore, the region of the parameter space to be explored must be carefully selected for the target EC tool. 
If the region is left too wide, then the search space becomes large and the search time, correspondingly, becomes long. If, on the other hand, the region is constrained too much, sub-optimal solutions may be found by \name.
Our analysis in Table~\ref{tab:simann} captures some interesting empirical results. For example, choosing a word length of 4 for LM training produces better results than any other word length. This observation is consistent across multiple datasets. "This motivates us to study in future work the effect of word length on LM training in greater detail."
%

\section*{Conclusion}
Configuration parameters play a crucial role in tuning EC tools. Manually selecting configuration parameters will degrade the quality of error correction (EC), and impact downstream genome assemblies~\cite{heydari2017evaluation}. \name automatically tunes these parameters without the need for a ground truth reference genome. 
Further, we improve our pipeline by leveraging character-level and word-level transformer networks. We observe that grouping nucleotides into groups of four ($|w| = 4$) in our LM causes the highest negative correlation with aligned reads, and gives the best results for both Lighter~\cite{song2014lighter} and LoRDEC~\cite{salmela2014lordec}, exemplar short- and long-read EC tools, respectively. In all cases, \name can configure the EC tool so that the alignment rate is what would be found by an exhaustive search of the parameter space or the alignment rate is within 1\% of the optimum. Further, \name has significant speed up of state-of-practice (actually doing the alignment, such as Bowtie2, over which we have a 80$\times$ -- 275$\times$ speedup) or the state-of-the-art autotuner for genomic reads, namely \textit{Athena} ($18\times$ speedup). 

Further, the LMs we have developed may potentially impact future work, unraveling the nuances of the language of the genome. The use of attention mechanisms opens up several avenues toward the interpretability of LM. The values of the attention weights can be analyzed; we expect that beyond a certain distance from a token, the weights will fall below a certain threshold. By finding and quantifying an appropriate threshold, we can reduce the work of the LM by preventing it from encoding information beyond a certain point. Furthermore, the patterns observed through attention may allow future work to explore and find biologically significant portions of the genome. This can play a critical role in the analysis of genetic mutations and their consequences.
\name can be applied to tune the configuration of many different EC tools. Due to its speed, it can be applied when datasets change or the instruments change, even transiently, to have different error characteristics. 

We now discuss a few aspects of \name, such as supporting EC tools, which are not $k$-mer based, the impact of sub-sampling on the perplexity metric, and the tuning complexity for the hyperparameters in \name.  

\noindent \textbf{Support for non $k$-mer based EC tools:} Although our evaluation  is performed on $k$-mer-based EC tools, such as Lighter and LoRDEC, the rationale for \name is not tightly coupled to $k$-mer-based correction techniques. Rather, it can be generalized to tune other parameters that impact the correction accuracy. This is achieved by \name's pipeline that treats the EC tool under consideration as a blackbox. For example, tools such as Racer~\cite{ilie2013racer}, which require users to specify the \textit{genome length} for the sequenced portion of the genome, can also be tuned with \name's pipeline by applying \name's searching heuristic over the space of possible \textit{genome lengths} (instead of the space of $k$-mer values). The selection criteria will be selecting the \textit{genome length} that minimizes the perplexity metric for the corrected reads. 

\noindent \textbf{Impact of Sub-sampling on Perplexity:} As shown in Figure~\ref{fig:overview}, \name performs sub-sampling over the original set of reads before feeding it to the EC tool being tuned. This is performed to reduce the searching time. However, \name's LM is trained over the 
\textit{entire} (uncorrected) dataset to exploit the high coverage of the full set of reads, which allows the LM to accurately capture the probability distribution of the sequence. 
The sub-sampling ratio is lower bounded so that the resulting coverage of the sample is $\geq$ 30$\times$, which is essential for the EC tool to perform efficient error correction~\cite{liu2013estimation}. 
Also notice that \name uses the perplexity scores that correspond to different $k$-values to pick the best among them. Accordingly, even if sub-sampling is impacting the perplexity scores, 
as long as the relative ordering of the perplexity scores with different $k$-values is maintained, \name can still select the best $k$-value.

\begin{figure}
    \centering
    \includegraphics[scale = 0.23]{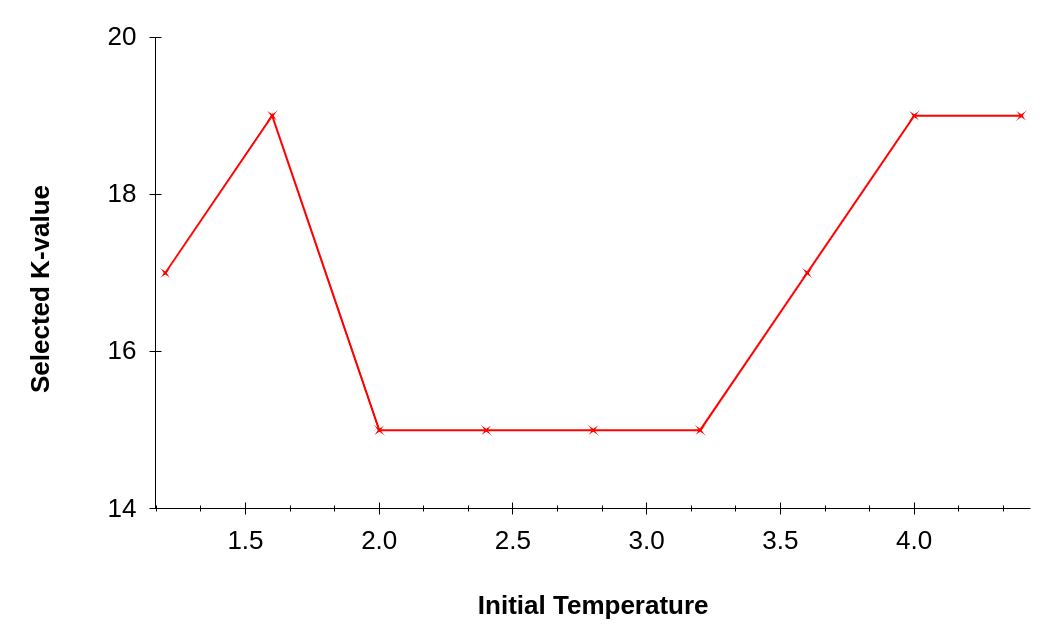}
    \caption{The selected value of $k$-mer length for Illumina dataset $D2$ [Table~\ref{tbl:Datasets_Description}]. The value of the parameter selected is relatively unaffected by the initial temperature ($T$). In the worst case, the alignment rate 61.15\% compared to 61.42\% at the optimum value. In this experiment, we set $\alpha$ = 0.7.}
    \label{fig:temp}
\end{figure}

\begin{figure}
    \centering
    \includegraphics[scale = 0.23]{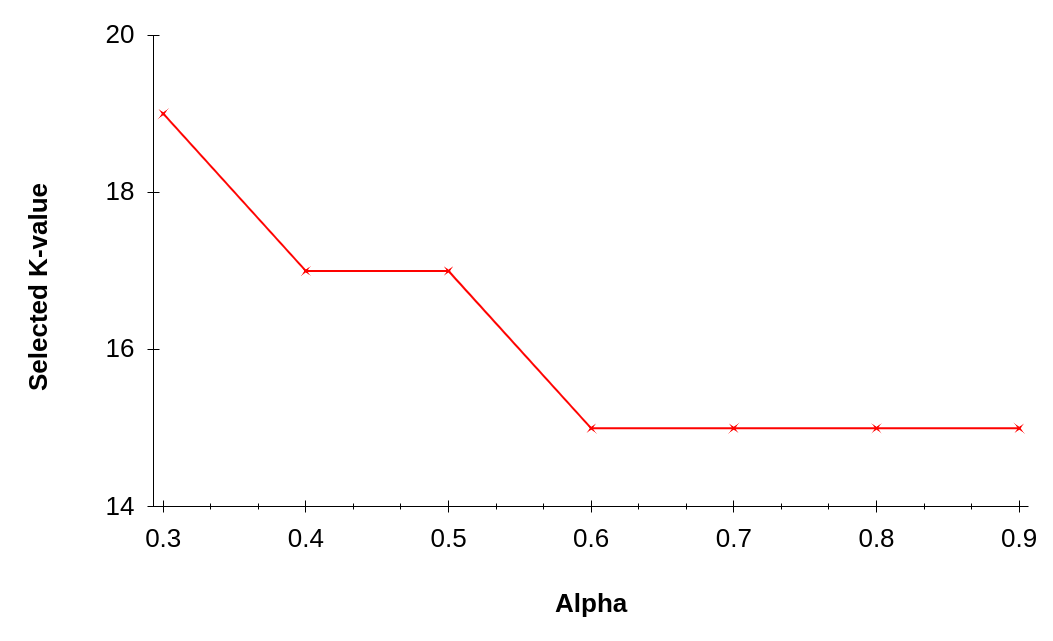}
    \caption{Variation of selected $k$ with parameter $\alpha$ keeping the initial temperature ($T$) = 2.8. In this case we observe that any $\alpha$ greater than 0.6 and less than 1 gives the optimum value for dataset $D2$.}
    \label{fig:alpha}
\end{figure}

\noindent \textbf{Hyperparameter tuning for \name's search heuristic:} \name relies on a Simulated Annealing (SA)-based searching to find the best $k$-value. However, SA has its own hyperparameters, such as initial temperature ($T$), temperature scale factor ($\alpha$), and number of cycles. It may appear that we have replaced one parameter selection for a different, yet equally hard, parameter selection problem. Fortunately, we find empirically that \name's performance is not sensitive to these parameters and a reasonable selection for their values (using common rules of thumb) provides the desirable performance. For example, a good rule of thumb for the ``Initial Temperature" is to pick a value that accepts about 98\% of the solutions to explore. As we can see from Fig.~\ref{fig:temp}, the selection made by \name stays constant for any value of temperature between 2.00 and 3.25, and even outside of that, the fall in alignment quality is not that high. Fig.~\ref{fig:alpha} shows that the final choice of \name is relatively unaffected by the other parameter of SA, $\alpha$. Moreover, prior works such as~\cite{ben2004computing} proposed simple algorithms that can automatically select appropriate values for SA's hyperparameters, which can be easily integrated in \name's pipeline. 

\noindent \textbf{Relevance of \name with newer genome sequencing tools:} As genomic reads become longer, the improvement due to \name will become more significant. This will happen because the longer-read technologies are more error prone, with PacBio and Oxford Nanopore reaching error rates of 15\% and 40\%~\cite{korlach2013understanding, laver2015assessing} respectively. Compare this to the error rates of less than 0.1\% with short reads~\cite{ma2019analysis}. 
Second, with longer reads, the possible range of $k$-values in $k$-mer-based EC tools will increase. This will mean that it will be more time consuming to perform an exhaustive search through the parameter search space. Hence, the efficient design of \name will be of greater benefit to the community.
\section*{Methods}

The proposed method for automatic tuning of EC Tools takes in the uncorrected reads and the EC Tool as the input, and involves the following 4 steps before generating the optimal input parameter ($k$) for the given EC Tool.
\begin{enumerate}
    \item \textbf{Sub-sampling}: The reads are usually obtained in the \textit{fasta} file format. We sample out a small portion of the dataset, which typically constitutes 4 to 5\% of the entire dataset. For example, for the dataset D7 (short reads, human genome), we sampled out 700k reads out of 15M reads to work on. We have empirically found that this fraction has good enough coverage for the entire pipeline to be reliable. The sub-sampling step helps us achieve a significant improvement in runtime.
    \item \textbf{LM training}: We explore both character as well as word level training for both short and long reads. We train the transformer network, that is, our language model on the uncorrected subsampled dataset and use the trained model for evaluating the corrected reads. 
    \item \textbf{Data correction}: Here we use the standard EC Tools to correct the uncorrected reads. We use Lighter for short reads and LoRDEC for long read correction. \revv{In principle, any EC tool can be used with \name.} The EC Tool generates different outputs for different input $k$ values and our aim is to find the $k$ that corresponds to the best correction.
    \item \textbf{Evaluation}: We take the corrected outputs and evaluate them using our trained language model. To accelerate the process of evaluation, we have made use of JIT compiler. Our metric of evaluation is perplexity that essentially captures how confident the learned model is during prediction. A lower perplexity corresponds to a higher confidence and vice-versa. We repeat steps 3 and 4 iterating with different $k$ values using a search technique called 'simulated annealing' to finally converge into the most optimal $k$ value. We verify that the output of the EC Tool for this $k$ value corresponds to the best alignment with the reference genome.
\end{enumerate}
The entire workflow is visually represented in Figure ~\ref{fig:overview}. 

We show the steps of our searching mechanism in algorithms \ref{algo1} and \ref{algo2}. A Transformer LM is trained on a given dataset. The range of $k$; $k_{min}$ and $k_{max}$, the LM, and the subsampled dataset $S'$ are given as input to the algorithm. We set the parameter $\delta$ (\ie, step size) = 1, the initial temperature $T_0 = 2.8$, and the constant $\alpha = 0.7$ ($\alpha$ represents the factor by which temperature is scaled after each cycle). These parameters are empirically found to perform the best across all datasets (including short and long reads) and therefore no additional tuning efforts are needed here. The search is allowed to run for 8 cycles, with 3 trials per cycle. 

\begin{algorithm}[h]
\caption{Correct Set of Reads}
\begin{flushleft}
{

\textbf{Input:} Dataset: $D_{0}$, Minimum value of $k$: $k_{min}$, Maximum value of $k$: $k_{max}$ \\
\textbf{Output:} Corrected Dataset: $D_c$ 
}
\end{flushleft}
\begin{flushleft}
\textbf{1-} Train a Transformer Language Model (LM) using $D_{0}$. \\
\textbf{2-} Select random sample $S'$ from $D_{0}$.\\
\textbf{3-} Call \textbf{Search $k$}($S'$, LM, $k_{min}$, $k_{max}$): \textbf{Algorithm \ref{algo2}}\\
\textbf{4-} Use the value of $k$ from step 3 and do a complete correction on the entire dataset $D_0$ and return $D_c$.

\end{flushleft}
\label{algo1}
\end{algorithm}

The probability of selecting a proposed solution that corresponds to a higher perplexity (a less likely sequence) is given by $p = \exp{(-\Delta E/(T\times\Delta E_{avg}))}$, where $E$, the energy variable, determines the volume of the search space. Larger $E$ values correspond to larger search space and viceversa. It is clear that as temperature $T$ is reduced, $p$ is reduced. Hence, the values assigned to parameters $T_0$ and $\alpha$ determine the overall performance of the search. If the temperature is reduced quickly over various cycles, by reducing $\alpha$, the search is restricted to a local neighborhood (exploitation). However, taking a value of $\alpha$ close to 1 causes the algorithm to explore a greater search space, and prevents it from converging to a local optimum.

\begin{algorithm} [h]
\caption{Search $k$}
\begin{flushleft}
{
\textbf{Input:} Dataset: $S'$, Language Model: LM, Minimum value of $k$: $k_{min}$, Maximum value of $k$: $k_{max}$  \\
\textbf{Output:} Best $ k: k^{*}$
}
\end{flushleft}


\begin{flushleft}
\textbf{1-} Randomly chose a value $k_c$ in range $(k_{min}, k_{max})$.\\
\textbf{2-} Initialize temperature T and constant $\alpha$ in range (0,1) \\
\textbf{3-} Initialize $\Delta E_{avg}$ = 0, $n_a = 0$ and constant $\delta$\\
\textbf{4-} Initialize the number of cycles and the number of trials in each cycle\\
\textbf{5-} for i in the range(0,cycles)\\
\textbf{6-} \quad for j in the range(0,trials)\\
\textbf{7-} \quad \quad Select random integer $k_r$ in range $(max(k_{min},k_c-\delta),min(k_{max},k_c+\delta))$\\
\textbf{8-} \quad \quad Calculate perplexity $f(LM,S',k_r)$ and $f(LM,S',k_c)$\\
\textbf{9-} \quad \quad Calculate $\Delta E$ = $|f(LM,S',k_r)-f(LM,S',k_c)|$ \\
\textbf{10-} \quad \quad if  $f(LM,S',k_r)>f(LM,S',k_c)$ Then:\\
\textbf{11-} \quad \quad \quad \quad if i=0 and j=0 Then:\\
\textbf{12-} \quad \quad \quad \quad \quad $\Delta E_{avg}$ = $\Delta E$\\
\textbf{13-} \quad \quad \quad \quad $p = \exp{(-\Delta E/(T\times \Delta E_{avg}))}$\\
\textbf{14-} \quad \quad \quad \quad Randomly select $\beta$ in range (0,1)\\
\textbf{15-} \quad \quad \quad \quad if $\beta < p$ Then:\\
\textbf{16-} \quad \quad \quad \quad \quad Set value of $k_c = k_r$\\
\textbf{17-} \quad \quad else\\
\textbf{18-} \quad \quad \quad \quad Set value of $k_c = k_r$\\
\textbf{19-}\quad \quad Increment $n_a$\\
\textbf{20-}\quad \quad Calculate $(\Delta E_{avg} \times (n_a-1.0) +  \Delta E) / n_a$\\
\textbf{21-}\quad \quad Update the value of $\Delta E_{avg}$ with value from step 20\\
\textbf{22-} \quad Update temperature T = $T \times \alpha$\\
\textbf{23-} return $k_c$
\end{flushleft}
\label{algo2}
\end{algorithm}

\section*{List of abbreviations}

JIT: Just in time \\
EC tool: Error correction tool \\ 
NLP: Natural language processing \\
LM: Language model \\
RNN: Recurrent neural network \\
ML: Machine learning \\
CLR: Continuous long reads \\
SA: Simulated annealing \\
\section*{Declarations}
\subsection*{Ethics approval and consent to participate}
Not applicable
\subsection*{Consent for publication}
Not applicable
\subsection*{Availability of data and materials}
Our code has been made available at a public github repository at: \url{https://github.com/icanforce/lerna-genomics}.

\subsection*{Competing interests}
No competing interests by Adobe Systems Inc. Prof. Chaterji is an Associate Editor of BMC Bioinformatics. 
\subsection*{Funding}
This work is supported in part by the NIH R01 Grant 1R01AI123037, a Lilly Endowment grant, a gift from Adobe Research, and funding from Purdue’s College of Engineering and Department of Agricultural and Biological Engineering. Any opinions, findings, and conclusions or recommendations expressed in this paper are those of the authors and do not necessarily reflect the views of the funding agencies.
\subsection*{Authors' contributions}
A.S., P.J., A.M., Z.Z., K.M., and S.C. participated in the computational analyses. A.S worked on the language model training and long reads analysis. P.J. worked on the short reads analysis, simulated annealing, and language model testing speedup using JIT compiler.  A.M. assisted with the EC tool usage, and Z.Z. evaluated the entire pipeline on real datasets reported in Tables~\ref{tab:real_pac} and ~\ref{tab:real_nano}. K.M. helped with finding the right datasets for the work. All authors wrote the manuscript. S.C. provided overall guidance and funding for the project. All authors read and edited the final manuscript.
\subsection*{Acknowledgments} 
None.

\begin{backmatter}




\bibliographystyle{bmc-mathphys} 
\bibliography{bmc_article}      
\end{backmatter}
\end{document}